\documentclass{svproc}
\usepackage[a4paper,margin=2cm]{geometry} 
\usepackage{url}
\usepackage{booktabs}
\usepackage{amsmath,amssymb}
\usepackage{graphicx}
\usepackage{subcaption}
\usepackage{float}

\begin{document}
\mainmatter              
\title{Understanding Friendship Formation with Explainable Machine Learning}
\titlerunning{Understanding friendship formation with explainable machine learning}  
%
\author{Mar\'ia Pereda\inst{1}\inst{2}
\authorrunning{Mar\'ia Pereda} 
%
%
\institute{Grupo de Investigación Ingenier\'ia de Organizaci\'on y Log\'istica (IOL), Departamento Ingenier\'ia de Organizaci\'on, Administraci\'on de empresas y Estad\'istica, Escuela T\'ecnica Superior de Ingenieros Industriales, Universidad Polit\'ecnica de Madrid, 28006 Madrid, Spain,\\
\email{maria.pereda@upm.es}\\ 
\and
Grupo Interdisciplinar de Sistemas Complejos (GISC), 28911 Legan\'es, Madrid, Spain}
}

\maketitle              

\begin{abstract}
Understanding the formation of social ties requires disentangling the roles of individual traits and local network structure. We analyse signed social relationships among 3,395 students using an interpretable machine learning model—the Explainable Boosting Machine (EBM)—to predict link polarity from individual attributes (prosociality, cognitive reflection, and gender) and a structural metric, triadic influence. Our results show that triadic influence overwhelmingly dominates link sign prediction, confirming that local network structure is the primary driver of social relationships. Nevertheless, a small subset of links (0.24\%) is primarily explained by individual-level traits. A detailed characterisation of this subset reveals that these links do not arise from distinct structural conditions, but rather correspond to weaker and less structurally embedded relationships. In particular, they are more likely to be negative ties and exhibit lower levels of structural balance, whereas triadic-dominant links are strongly associated with positive relationships and highly balanced configurations. Furthermore, we find that the signs of links without indirect structural paths are not explained by individual traits, but by the absence of structural reinforcement itself. Taken together, these findings support a layered view of social tie formation, in which structural mechanisms dominate globally, while individual-level effects emerge in specific, less constrained contexts. More broadly, our work highlights the value of explainable machine learning for uncovering the mechanisms underlying social network formation.

\end{abstract}
\section{Introduction}

Social networks shape human interactions from early childhood to adult life, but the mechanisms behind the formation of individual social ties remain elusive. In particular, understanding whether friendships emerge primarily from intrinsic individual traits or from the social environment is a longstanding question in network science.

In a recent study \cite{ruiz2023triadic}, it was shown that the sign of a relationship between two students—friendship or enmity—could be accurately predicted using a simple structural metric called \emph{triadic influence}. This measure quantifies the effect of indirect paths of length two between two students, capturing the mediating role of mutual contacts. Interestingly, triadic influence alone outperformed models that included personal traits such as gender, cognitive reflection, or prosociality. These findings suggest that local structural features can act as a proxy for deeper compatibility traits that may be otherwise difficult to measure directly.

Despite the strong performance of triadic influence, the same study observed that in the absence of such structural paths—that is, for isolated links without directed paths of length two—the prosociality of both students still yielded predictions that were better than random. This points to a plausible scenario where prosociality drives the nucleation of early links, which are later embedded in growing triadic structures that take over as dominant mechanisms of network formation.

In this work, we revisit this hypothesis from a different perspective. Rather than using black-box models like deep neural networks, our goal is to interpret link formation in these networks by identifying which features—individual or structural—most strongly influence it. To this end, we perform a systematic comparison of machine learning classifiers, selecting among them the Explainable Boosting Machine (EBM) for its combination of strong predictive performance and interpretability. EBMs are a type of Generalized Additive Model that learns transparent feature-wise effects, making them ideal for uncovering the roles of individual traits and local topology in a controlled and explainable manner.

Our analysis confirms that: 
(I) triadic influence is the dominant predictor of link sign across the network; 
(II) a small subset of links is primarily explained by individual-level traits rather than structural effects; 
(III) these links do not arise from fundamentally different structural conditions, but tend to be less strongly embedded and more frequently associated with weak or negative ties; 
(IV) contrary to previous interpretations, links without indirect structural paths ($A^2_{ij} = 0$) are not explained by individual traits, but by the absence of structural reinforcement itself.

Through the lens of explainable machine learning, we provide new insight into the mechanisms of friendship formation in adolescent networks, and offer evidence consistent with a two-stage dynamic: early links forming under different conditions—potentially before structural patterns stabilize—and later links being more predictable via structural mechanisms such as triadic closure.

\section{Methods}

\subsection*{Dataset and Features}

We use publicly available data from a previous study~\cite{ruiz2023triadic}, comprising self-reported signed social relationships among 3,395 students from 13 secondary schools in Spain. Each directed link represents a declared relationship from student $i$ to student $j$, with ratings categorized into four levels (from $-2$: “very bad” to $+2$: “very good”). For binary classification, we focus on links labeled as either friendship ($+2$ or $+1$) or enmity ($-2$ or $-1$).

Each student is also characterized by three individual traits:
\begin{itemize}
    \item \textit{Gender}, self-reported (male, female, non-binary).
    \item \textit{Cognitive reflection}, assessed via the standard three-question Cognitive Reflection Test (CRT) \cite{Branas-Garza2019CognitiveWhen}. The CRT captures the tendency to override an intuitive but incorrect answer in favour of a more deliberate response. The score ranges from 0 to 3, corresponding to the number of correct responses.
    \item \textit{Prosociality}, operationalised through a set of three economic choice questions inspired by experimental paradigms on distributional preferences \cite{Fehr2008EgalitarianismChildren} \cite{AlfonsoCostillo2022TheTeenagers}, scaled to the interval $[0,1]$.
\end{itemize}

In addition to individual traits, we include a network-based feature: \textit{triadic influence}~\cite{ruiz2023triadic}, defined as the sum over all directed paths of length 2 from $i$ to $j$:  
\[
I_{ij} = \sum_k w_{ik} w_{kj}.
\]
This metric captures the indirect social context influencing the $i \rightarrow j$ link and has been shown to strongly correlate with link polarity.

\subsection*{Classification and Model Selection}

Our task consists in predicting the sign of a directed relationship (link sign prediction) between two students. Input features include individual traits for both students and/or the triadic influence between them. We compare four classifiers: two deep neural networks (TensorFlow, PyTorch), XGBoost, and the Explainable Boosting Machine (EBM).

All models are optimised using stratified nested cross-validation (5 outer and 5 inner folds)~\cite{Anderssen2006ReducingValidation}~\cite{Varma2006BiasSelection} to ensure robustness and unbiased performance estimation. To address the class imbalance present in the dataset, we incorporated Random OverSampling within the training pipeline. This oversampling step is applied exclusively to the training folds during each iteration of the nested cross-validation, thereby preserving the integrity of the validation folds and preventing data leakage. Embedding oversampling within the pipeline ensures that the resampling procedure is integrated seamlessly with feature scaling and model fitting, promoting a more robust and unbiased estimation of the model’s generalisation performance. In addition, all feature variables were standardised using \texttt{StandardScaler} to ensure zero mean and unit variance, as required by several machine learning algorithms (e.g., distance-based and regularised models). Due to class imbalance, we report balanced accuracy (mean of true positive and true negative rates) as our main evaluation metric (table \ref{tabla}).

\begin{table}[t]
\centering
\caption{Balanced accuracy and standard deviation (SD) across models.}
\label{ModelSelection}
\begin{tabular}{lccc}
\toprule
Model & Accuracy & SD & Key Hyperparameters \\
\midrule
TF Neural Network & 0.824 & 0.005 & hidden\_units=32, lr=0.001, batch=32 \\
PyTorch Neural Net & 0.824 & 0.005 & hidden\_units=16, lr=0.05, batch=32 \\
XGBoost & 0.824 & 0.004 & depth=3, estimators=70, lr=0.1 \\
EBM & 0.824 & 0.002 & interactions=5, bins=128 \\
\bottomrule
\end{tabular}
\label{tabla}
\end{table}

While all models perform equally well in terms of accuracy, the EBM yields the lowest variance, suggesting more stable predictions. Its transparent additive structure also enables direct interpretation of feature contributions, making it the model of choice for subsequent analysis.

\subsection*{Explainable Boosting Machine}

The Explainable Boosting Machine (EBM) is a type of machine learning model designed to be both accurate and interpretable \cite{Caruana2015IntelligibleHealthCare}\cite{Nori2019InterpretML}, making it especially useful when we want to understand not just \textit{what} the model predicts, but \textit{why}. It is inspired by Generalised Additive Models (GAMs)~\cite{Hastie1990GeneralizedModels}, which assume the response variable $Y$ follows a distribution from the exponential family. The model structure is given by:

\begin{equation}
g(\mathbb{E}[Y]) = \beta_0 + \sum_j f_j(x_j),
\label{eq:gam}
\end{equation}

where $g$ is a link function that allows the model to flexibly adapt to different problem types—typically regression or classification. The functions $f_j$ are univariate functions that capture the (potentially nonlinear) contribution of each predictor $x_j$ to the outcome.

In classification settings, the response variable $Y$ is categorical (e.g., $Y \in \{0,1\}$ in binary classification). To remain within the GAM framework, it is assumed that $Y$ follows a binomial distribution (in the binary case) or a multinomial distribution (in the multiclass case), both of which are members of the exponential family. A suitable link function $g$ is used to relate the expected value $\mathbb{E}[Y]$ to the predictors. For instance, in binary classification, one typically models:

\begin{equation}
Y \sim \mathrm{Bernoulli}(p), \quad \text{where } p = \mathbb{P}(Y=1 \mid X),
\end{equation}

and the link function is the logit function:

\begin{equation}
\log\left(\frac{p}{1 - p}\right) = \beta_0 + \sum_j f_j(x_j) + \sum_{i \neq j} f_{ij}(x_i, x_j),
\label{eq:logit}
\end{equation}

so that the model estimates the log-odds, which are then transformed into probabilities via the logistic function.

The EBM extends the traditional GAM by including explicitly modelled pairwise interactions, leading to a formulation of the form:

\begin{equation}
g(\mathbb{E}[Y]) = \beta_0 + \sum_j f_j(x_j) + \sum_{i \neq j} f_{ij}(x_i, x_j),
\label{eq:ebm}
\end{equation}

where $f_{ij}$ denotes the interaction function between variables $x_i$ and $x_j$. These interaction terms are identified through the GA$^2$M algorithm~\cite{Lou2013AccurateInteractions}, which performs a feature selection process based on statistical significance, followed by a forward selection strategy that retains only relevant interactions.

Unlike traditional GAMs, the EBM fits the $f_j$ and $f_{ij}$ functions using gradient boosting with ensembles of shallow regression trees. This is done through a round-robin process with a low learning rate and many boosting iterations, which ensures stability and robustness, particularly in the presence of correlated predictors~\cite{Nori2019InterpretML}. This iterative approach mitigates feature dominance due to ordering and collinearity.

Despite the use of boosting, EBMs remain interpretable. Each term—whether univariate or interaction—can be visualised, enabling a transparent understanding of how the prediction is formed. In particular, the model provides partial dependence plots for individual predictors and two-dimensional plots for interaction effects. These features allow for a detailed interpretation of the model's internal logic and a transparent view of the variables that influence system behaviour, both in isolation and in interaction. This is especially relevant in contexts such as social dynamics, where identifying the factors driving specific relational outcomes is of substantive interest.

\section{Results}
\subsection{Triadic influence is the dominant predictor of social links}

To assess the overall contribution of each variable to the prediction of link sign, we examined the global explanation produced by the Explainable Boosting Machine (EBM). This global plot ranks all features according to their average importance across the dataset, as quantified by the mean absolute contribution of each term to the model's output log-odds. This metric reflects how strongly each feature influences the predicted probability of a positive link, aggregated over all student pairs.

\begin{figure}[t]
    \centering
    \includegraphics[width=0.9\textwidth]{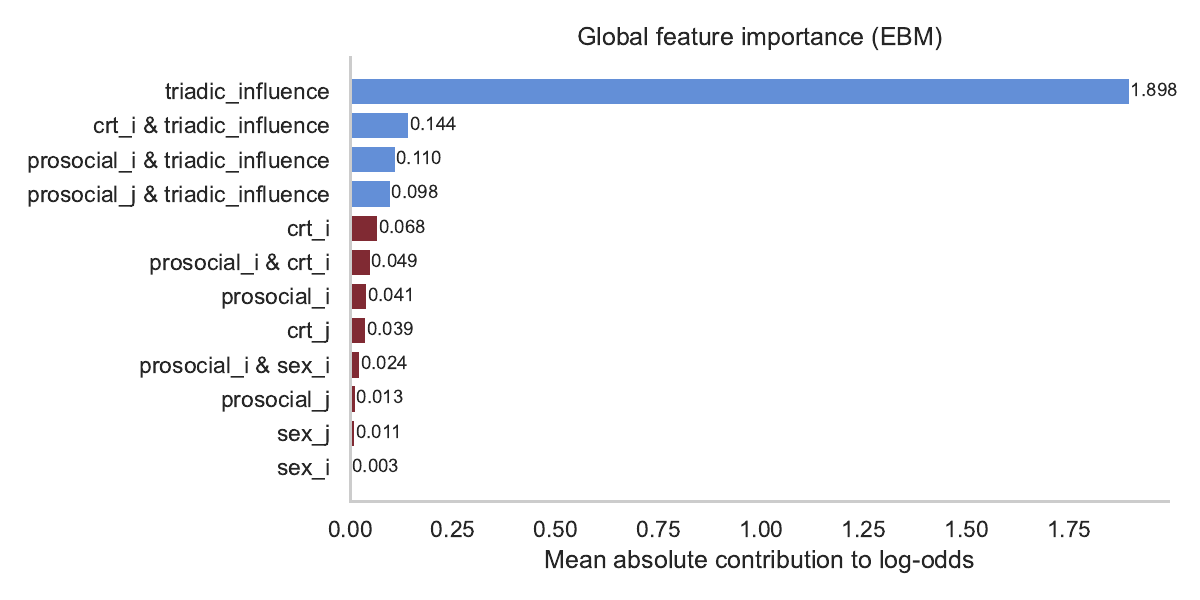}
    \caption{Global feature importance derived from the EBM. Each bar represents the average contribution of the corresponding predictor to the model's log-odds output, across all samples. Triadic influence is by far the most influential variable.}
    \label{fig:global_importance}
\end{figure}

Figure~\ref{fig:global_importance} shows the resulting ranking of feature importance derived from the Explainable Boosting Machine (EBM). The presented plot illustrates the relative impact of each feature on the model's predictions by showing their mean absolute contributions to the log-odds of a positive link. In this context, the log-odds represent the natural logarithm of the odds ratio, where the odds ratio quantifies the likelihood of a positive link occurring versus not occurring. Expressing contributions in terms of log-odds enables a linear additive interpretation of how each feature shifts the model's predicted probability on a logarithmic scale. This approach is particularly appropriate for classification models based on logistic regression or related frameworks, such as the EBM used here.

Among all predictors, \textit{triadic influence} emerges as the dominant factor, far surpassing any individual-level characteristic such as prosociality, cognitive reflection, or gender. Our analysis reveals that 91.47\% of the examples are dominated by triadic influence contributions exceeding 50\% of the total feature impact, whereas only 0.24\% are dominated by personal characteristics, and 1.27\% by interaction features—namely, pairwise combinations of personal traits and triadic influence. The remaining 7.02\% of examples do not show a clear dominance by any category, reflecting cases where influence is more evenly distributed across features.

Thus, larger absolute contributions correspond to stronger influences on the predicted probability of link formation. Features with higher importance values induce more substantial changes in the log-odds, thereby playing a greater role in determining the presence or absence of social ties in the network. This quantitative representation facilitates the identification of key structural and individual factors shaping the prediction of link sign.

This result confirms previous findings that local structural information, specifically the presence and alignment of directed two-step paths, plays a central role in shaping social interactions in signed networks. In contrast, individual attributes such as prosociality or cognitive style contribute significantly less to the model's overall behaviour. This suggests that while these traits may explain particular subsets of links (see Section~\ref{sec:Personal}), they do not account for the majority of the predictive structure in the data. The observed dominance of triadic influence underscores the explanatory power of mesoscopic network topology in understanding the formation of social bonds.

\subsection{Interpreting feature effects through dependence plots}

While global feature importance provides a ranking of predictors, it does not reveal how each variable influences the model’s output. To gain a more detailed understanding of the mechanisms underlying link formation, we examine the feature-wise functions learned by the Explainable Boosting Machine. These functions, commonly visualised through dependence plots, represent the marginal contribution of each predictor to the model’s log-odds as a function of its value.

In the EBM framework, each variable contributes additively to the prediction through a function $f_j(x_j)$ (Eq.~\ref{eq:ebm}). Dependence plots therefore provide a direct visualisation of these learned functions, allowing us to interpret how changes in a given feature affect the probability of a positive link. Unlike traditional partial dependence plots, these functions are intrinsic to the model and can be interpreted without additional approximation.

We analyse these effects variable by variable, focusing on both structural and individual predictors. This approach allows us to move beyond global importance and uncover the functional form of each contribution, identifying whether effects are monotonic, nonlinear, or localised to specific ranges of the feature space.

\begin{figure}[ht]
    \centering
    \includegraphics[width=0.6\textwidth]{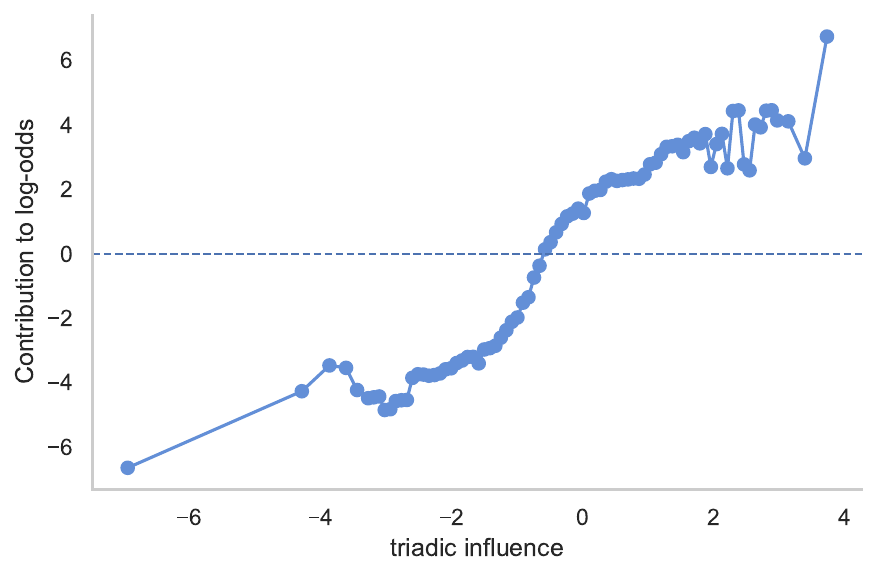}
    \caption{Dependence plot for triadic influence. The curve represents the contribution of triadic influence to the model’s log-odds output as a function of its value. Positive values increase the probability of friendship, while negative values favour enmity.}
    \label{fig:triadic_dependence}
\end{figure}

We begin with \textit{triadic influence}, the dominant predictor identified in the previous section. Its dependence plot reveals a strongly structured and approximately monotonic relationship between the value of triadic influence and the model’s output (Figure~\ref{fig:triadic_dependence}). As the value of triadic influence increases, the contribution to the log-odds becomes progressively more positive, indicating a higher probability of a positive link. Conversely, negative values of triadic influence lead to increasingly negative contributions, associated with a higher likelihood of enmity. The transition between these regimes is smooth and centred around zero, suggesting that the absence of indirect reinforcement does not introduce additional effects beyond the neutral baseline. This pattern is consistent with the interpretation of triadic influence as a measure of structural reinforcement: aligned indirect paths promote positive ties, whereas conflicting paths favour negative ones. Importantly, the smooth and continuous nature of the curve indicates that this effect operates across the full range of the feature, rather than being driven by isolated thresholds or extreme values. This provides further evidence that local network structure acts as a primary organising principle in the formation of social ties.

\begin{figure}[h]
    \centering
    \begin{subfigure}{0.48\textwidth}
        \centering
        \includegraphics[width=\textwidth]{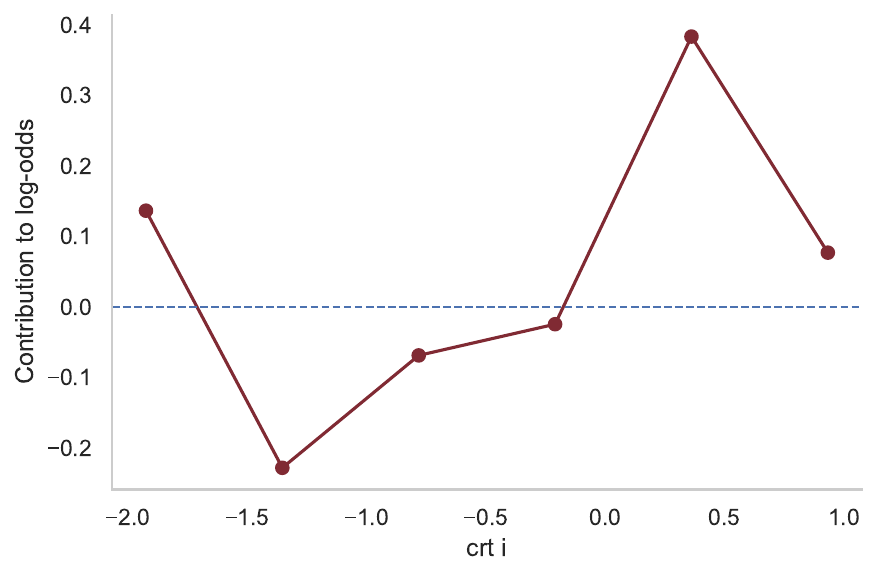}
        \caption{CRT of the source node ($i$).}
        \label{fig:crt_i}
    \end{subfigure}
    \hfill
    \begin{subfigure}{0.48\textwidth}
        \centering
        \includegraphics[width=\textwidth]{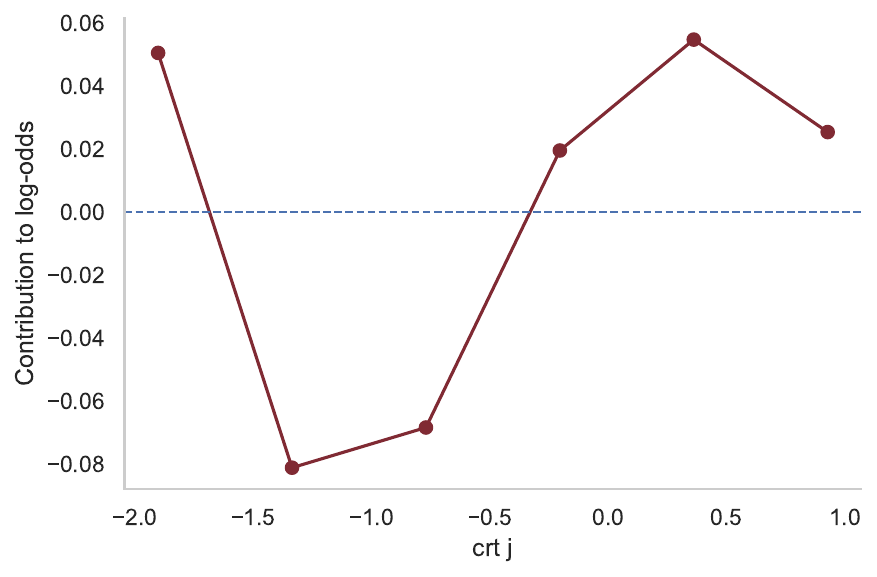}
        \caption{CRT of the target node ($j$).}
        \label{fig:crt_j}
    \end{subfigure}
    \caption{Dependence plots for cognitive reflection (CRT). Each panel shows the contribution of the variable to the model’s log-odds output.}
    \label{fig:crt_dependence}
\end{figure}

We next examine the effect of cognitive reflection, measured through the Cognitive Reflection Test (CRT). In our framework, each link involves two individuals, denoted as $i$ (the source of the directed relationship) and $j$ (the target). Accordingly, variables with subscript $_i$ refer to the characteristics of the individual initiating the tie, whereas $_j$ corresponds to those of the individual receiving it. Figure~\ref{fig:crt_dependence} shows the dependence plots for CRT scores of both individuals. In contrast to triadic influence, the effect of cognitive reflection is considerably weaker and less structured. The contributions remain close to zero across most of the range, indicating a limited global impact on sign prediction.

For the source node ($i$), higher CRT scores tend to be associated with a slight decrease in the probability of a positive link, although the effect is modest and not strictly monotonic. For the target node ($j$), the pattern is even weaker, with contributions fluctuating around zero and no clear directional trend. Overall, these results suggest that cognitive reflection does not act as a primary driver of link formation, but may exert a secondary and context-dependent influence. This stands in contrast with the strong and consistent effect observed for triadic influence, reinforcing the idea that individual traits contribute only marginally to the overall predictive structure of the network.

\begin{figure}[h]
    \centering
    \begin{subfigure}{0.48\textwidth}
        \centering
        \includegraphics[width=\textwidth]{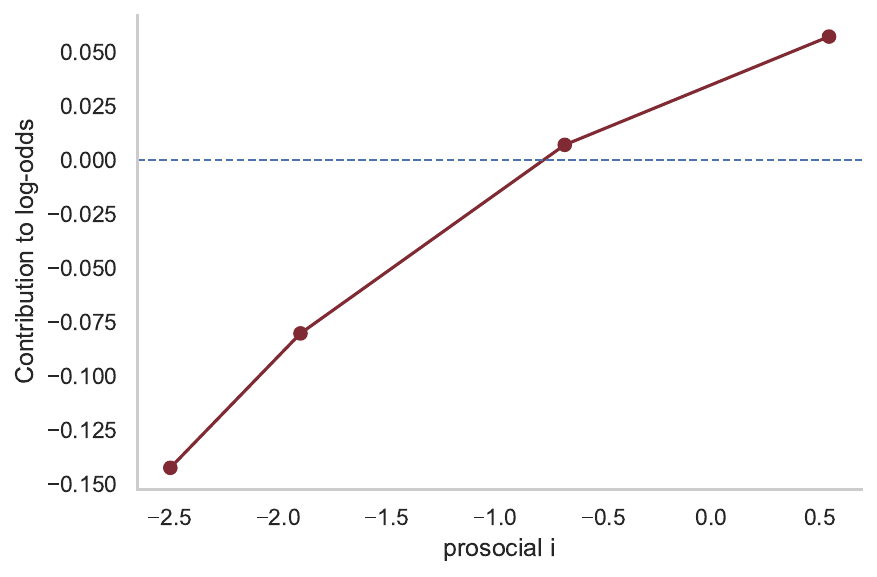}
        \caption{Prosociality of the source node ($i$).}
        \label{fig:prosocial_i}
    \end{subfigure}
    \hfill
    \begin{subfigure}{0.48\textwidth}
        \centering
        \includegraphics[width=\textwidth]{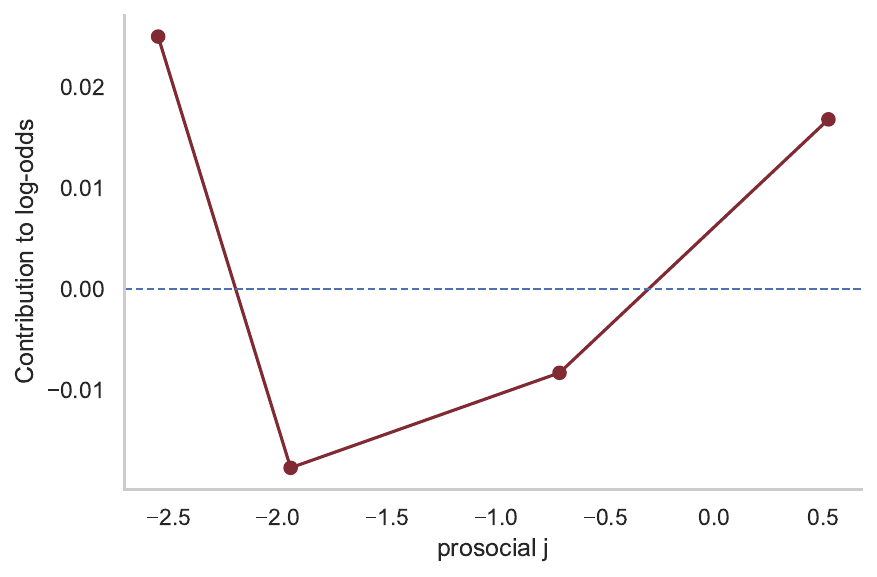}
        \caption{Prosociality of the target node ($j$).}
        \label{fig:prosocial_j}
    \end{subfigure}
    \caption{Dependence plots for prosociality. Each panel shows the contribution of the variable to the model’s log-odds output.}
    \label{fig:prosocial_dependence}
\end{figure}

We now turn to prosociality, which captures individual preferences for equitable and cooperative behaviour. As before, subscripts $_i$ and $_j$ denote the source and target nodes of the directed relationship, respectively. Figure~\ref{fig:prosocial_dependence} shows the dependence plots for prosociality. Compared to cognitive reflection, prosociality exhibits a slightly more structured effect, particularly for the source node. Higher levels of prosociality for individual $i$ are associated with an increase in the probability of a positive link, as reflected by progressively more positive contributions to the log-odds. Although the magnitude of this effect remains modest, the trend is relatively consistent across the range of values.

In contrast, the effect of prosociality for the target node ($j$) is considerably weaker, with contributions remaining close to zero and no clear monotonic pattern. This asymmetry suggests that prosocial behaviour primarily influences the formation of outgoing ties, rather than how individuals are perceived by others. In general, these results indicate that prosociality has a measurable but limited influence on link formation. Its effect is more pronounced than that of cognitive reflection, yet still substantially weaker and more localised than the structural signal captured by triadic influence. This reinforces the view that individual traits play a secondary role, modulating specific cases rather than driving the global organisation of the network.

\subsection{Individual-Level Dominance in a Subset of Links}\label{sec:Personal}

To characterise the mechanisms driving individual predictions, we analyse the relative contribution of different groups of features within the Explainable Boosting Machine (EBM). Specifically, we partition the predictors into three categories: individual-level traits (prosociality, cognitive reflection, and gender for both nodes), triadic influence, and interaction terms. For each link, we compute the sum of the absolute contributions of features within each category and define a category as dominant if it accounts for more than 50\% of the total contribution to the model’s log-odds. This criterion ensures that dominance reflects a clear majority influence rather than marginal differences between predictors.

Under this definition, we find that the vast majority of social links are dominated by triadic influence. In contrast, only a small subset of links (145 relationships, corresponding to 0.24\% of all examples) is dominated by individual-level characteristics such as prosociality and cognitive reflection. This minority indicates that, despite the overwhelming importance of network structure, alternative mechanisms of link formation do emerge in specific cases.

\begin{figure}[h]
    \centering
    \includegraphics[width=0.9\textwidth]{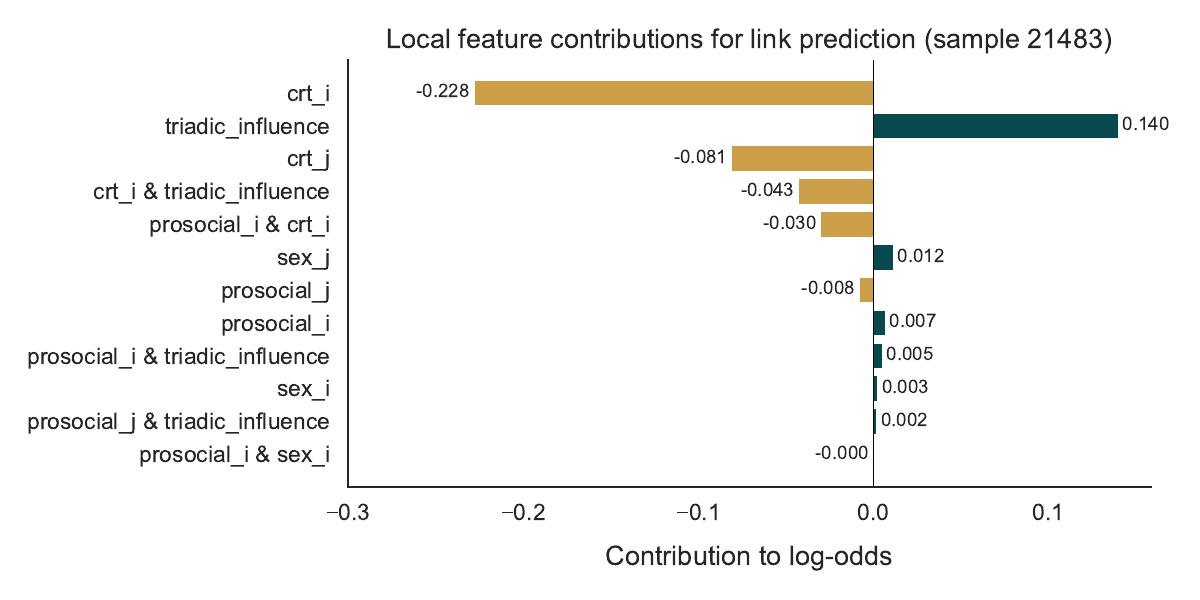}
    \caption{Local feature contributions for a single link sign prediction. Each bar represents the contribution of a predictor to the model's log-odds. Positive values increase the probability of a positive link, while negative values decrease it. In this example, the prediction is primarily driven by individual-level traits rather than structural information.}
    \label{fig:local_contributions}
\end{figure}

To illustrate this behaviour, Figure~\ref{fig:local_contributions} presents a representative local explanation for one such link. In this instance, individual characteristics outweigh structural effects in determining the predicted relationship. Cognitive reflection for the source node ($\texttt{crt\_i}$) provides the largest negative contribution, substantially reducing the probability of a positive link. Additional contributions from $\texttt{crt\_j}$ and interaction terms reinforce this effect. Although triadic influence contributes positively, its magnitude is insufficient to dominate the prediction.

This example illustrates that, while triadic influence governs the majority of links, individual-level traits can become the primary drivers in specific instances. Importantly, these cases do not appear to be random fluctuations, but rather reflect a distinct regime of link formation in which personal attributes outweigh structural reinforcement.

The existence of this subset highlights the heterogeneous nature of social tie formation. Rather than being governed by a single mechanism, the network exhibits a dual behaviour: a dominant structural regime driven by triadic closure, and a secondary regime in which individual predispositions play a decisive role at the level of specific links.

\subsection{Characterisation of individual-level dominant relationships}

To characterise the subset of links primarily predicted by individual-level attributes, we compare the distributions of relevant structural variables across two groups: links dominated by individual-level traits and links dominated by triadic influence.

Structural variables include the number of directed paths of length two between individuals, as well as the number of weak and strong connections of each type (enemy, dislike, like, and friendship). We define \textit{weak connections} as dyads with no directed paths of length two between them (i.e., $A^2_{ij} = 0$), indicating the absence of indirect structural reinforcement. In contrast, \textit{strong connections} correspond to dyads with at least one such path ($A^2_{ij} > 0$), reflecting the presence of triadic structure. In addition, we examine individual attributes such as prosociality and cognitive reflection (CRT) for both members of each dyad. Across these variables, we do not observe substantial differences between the two subsets. In particular, the distributions of weak connections are nearly identical, with the vast majority of dyads exhibiting no weak ties of any type. Similarly, the distributions of strong connections by type remain broadly comparable across both groups. Some differences emerge when considering the overall number of directed paths of length two. Links dominated by triadic influence tend to exhibit a broader distribution, with a heavier tail towards higher values, indicating a richer structural context. In contrast, links dominated by individual-level traits are more concentrated at lower values, suggesting comparatively weaker structural embedding. However, these differences are gradual rather than sharp, and do not define a clear boundary between the two subsets. Taken together, these results indicate that links dominated by individual-level traits do not correspond to a distinct structural regime. Instead, they arise within largely similar structural conditions, with only subtle differences in the intensity of indirect connectivity.

We next examine the distribution of individual-level attributes across both subsets, focusing on prosociality, cognitive reflection (CRT), and gender. Prosociality exhibits a mild but consistent shift between the two groups. Links dominated by individual-level traits tend to involve individuals with higher prosociality scores, with a greater concentration of mass at the upper end of the distribution. In contrast, links dominated by triadic influence are more concentrated around intermediate values. This suggests that prosocial behaviour may play a more prominent role in those cases where individual-level attributes dominate the prediction. By comparison, cognitive reflection shows largely similar distributions across both subsets, with only minor differences that do not indicate a clear separation. Likewise, gender appears uniformly distributed in both groups, with no observable imbalance. Taken together, these results indicate that, among the individual attributes considered, only prosociality exhibits a weak but systematic association with individual-level dominance. However, this effect remains limited in magnitude and does not define a sharply distinct subset, reinforcing the view that these links emerge within broadly similar individual and structural conditions.

We next examine the academic course year of the students involved in each dyad. Unlike the previous variables, this feature reveals a more noticeable shift between the two subsets. Links dominated by individual-level traits show a clear concentration in second-year students, with a substantially higher proportion of dyads involving this group. In contrast, links dominated by triadic influence are more evenly distributed across the early academic years, with comparable proportions of first- and second-year students and a smoother decay across higher courses. This pattern suggests that the balance between structural and individual drivers of link formation may evolve with time spent in the social environment. While structural mechanisms are already active from early stages, individual-level dominance appears to be more concentrated at specific points in the academic trajectory, rather than being uniformly distributed.

We further examine whether the nodes involved in each subset differ in their structural role within the network, as measured by standard centrality metrics, including degree, betweenness, and closeness. Across all measures, we do not observe meaningful differences between links dominated by individual-level traits and those dominated by triadic influence. The distributions of degree are nearly identical in both subsets, indicating that nodes involved in individual-level dominant links are neither more nor less connected than those in structurally driven links. Similarly, betweenness and closeness centralities show substantial overlap, with only minor variations that do not suggest a systematic distinction. These results indicate that individual-level dominance is not associated with specific types of nodes in the network. Instead, it emerges independently of node centrality, reinforcing the view that the distinction between the two subsets arises from how features combine at the level of individual links, rather than from the global position of the nodes involved.

\begin{figure}[h]
    \centering
    \includegraphics[width=0.4\textwidth]{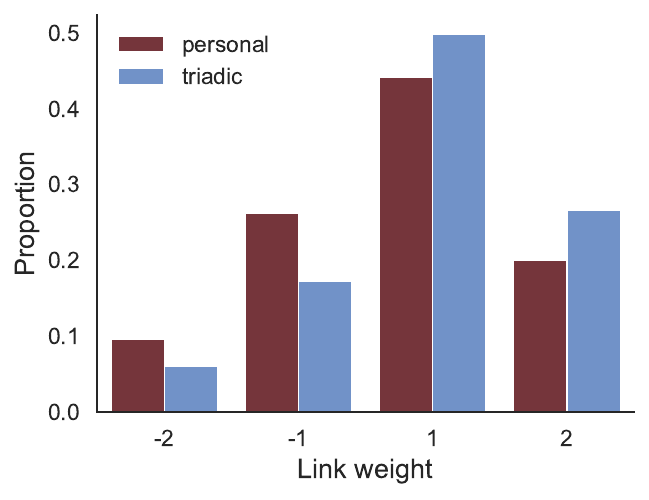}
    \caption{Distribution of link weights for edges dominated by individual-level traits (personal) and triadic influence. Structural dominance is associated with a higher proportion of strongly positive relationships, whereas individual-level dominance appears more frequently in weaker or negative ties.}
    \label{fig:link_weight}
\end{figure}

We next examine properties at the level of individual links, focusing on edge betweenness and link weight. Edge betweenness does not reveal substantial differences between the two subsets. The distributions largely overlap, with only a slightly heavier tail in the triadic-dominant group, suggesting that structurally driven links may occasionally occupy more central positions in the network. However, this effect remains limited and does not indicate a clear structural separation between the two types of links. In contrast, link weight exhibits a clear and systematic difference. As shown in Figure~\ref{fig:link_weight}, links dominated by triadic influence are more frequently associated with strongly positive relationships, with a higher proportion of high-weight (friendship) edges. Conversely, links dominated by individual-level traits display a more heterogeneous distribution, including a greater presence of weak and negative ties. This difference is quantitatively reflected in the proportion of negative links: 35.9\% of personal-dominant links are negative, compared to 23.5\% in the triadic-dominant subset. A chi-squared test confirms that this difference is statistically significant ($p = 0.00062$), with an odds ratio of 1.82, indicating that links dominated by individual-level traits are substantially more likely to be negative. Taken together, these results suggest that structural mechanisms are more closely associated with reinforced and stable social relationships, while individual-level effects tend to emerge in links that are less consolidated or more heterogeneous in nature.

\begin{figure}[h]
    \centering
    \includegraphics[width=0.9\textwidth]{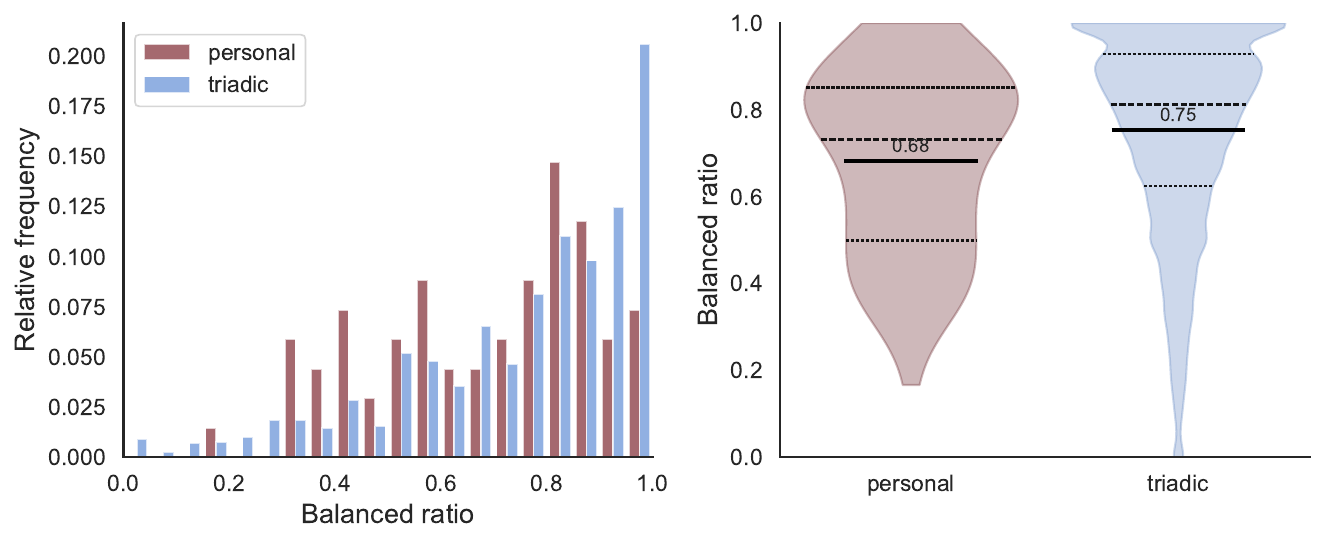}
    \caption{Balanced ratio distributions for links dominated by individual-level traits (personal) and triadic influence. (a) Relative frequency histogram showing the distribution of balanced ratios. (b) Violin plot showing the full distribution, quartiles, and mean values for each group.}
    \label{fig:balance}
\end{figure}

To further characterise the structural context of link formation, we analyse the data through the lens of balance theory in signed networks. This framework posits that relationships embedded in triads tend to evolve towards \emph{balanced configurations}, in which the product of the signs of the three edges is positive. For each link $(i,j)$, we identify all common neighbours $k$ that form triangles with it. For every such triangle $(i,j,k)$, we evaluate whether it is balanced by considering the signs of its three edges. Specifically, a triangle is classified as balanced if the product of the signs satisfies $s_{ij} \cdot s_{ik} \cdot s_{jk} > 0$, where $s_{uv} \in \{-1, +1\}$ denotes the sign of the relationship between nodes $u$ and $v$. We then compute, for each link, the fraction of these triangles that satisfy this condition. This quantity, referred to as the \emph{balanced ratio}, captures the extent to which a link is embedded in structurally consistent local configurations. Figure~\ref{fig:balance} compares the distribution of balanced ratios for links dominated by individual-level traits and by triadic influence. As shown in Figure~\ref{fig:balance}(b), triadic-dominant links exhibit higher balanced ratios on average (mean $\approx 0.75$) than personal-dominant links (mean $\approx 0.68$), indicating a stronger alignment with structurally coherent triads. Beyond differences in mean values, the shape of the distributions further reinforces this distinction. As shown in Figure~\ref{fig:balance}(a), the relative frequency distribution reveals a pronounced concentration of triadic-dominant links near one, indicating that they are frequently embedded in fully or nearly fully balanced configurations. In contrast, personal-dominant links display a broader distribution, with greater density at intermediate values, reflecting a more heterogeneous and less structurally constrained embedding. These results provide a structural interpretation of the differences observed throughout the paper. While triadic influence captures links that are embedded in coherent and socially reinforced configurations, individual-level dominance appears in links that are less structurally aligned, and therefore less constrained by balance dynamics.

\subsection{Links without structural paths are still explained by triadic influence}

Previous work~\cite{ruiz2023triadic} hypothesised that in the absence of structural connections, specifically, when there are no two-direct paths between individuals, link formation may be explained primarily by individual-level traits. To test this claim, we examined the subset of links in our dataset for which the number of directed paths of length two is zero, i.e., links for which $A^2_{ij} = 0$.

For each of these instances, we computed local explanations using the Explainable Boosting Machine (EBM), quantifying the contribution of each feature to the predicted sign of the link. Features were grouped into two categories: individual-level traits (prosociality, cognitive reflection test scores, and gender for both nodes in the dyad), and triadic influence.

Despite the absence of structural paths, we found that \emph{none} of the 1,337 links in this subset were dominantly explained by individual-level traits. In every case, the prediction was driven primarily by the triadic influence feature, whose value is zero in this subset by construction. That is, although the structural path count is zero, the model still interprets the absence of triadic reinforcement as informative, and no instance in this group is attributed to personal characteristics.

This finding suggests that, in our data, links without apparent structural context are not necessarily driven by latent personal affinities. Rather, the model captures their emergence (or lack thereof) through the absence of structural reinforcement itself, encoded via the triadic influence signal.

\section{Conclusions}

In this work, we have used an interpretable machine learning framework to revisit a central question in network science: to what extent are social ties driven by individual traits versus local structural mechanisms. By leveraging the Explainable Boosting Machine (EBM), we have been able not only to predict link polarity with high accuracy, but also to disentangle the contributions of different factors at both the global and local levels.

Our results provide clear evidence that triadic influence is the dominant organising principle of social relationships. Across the dataset, the vast majority of links are explained by structural reinforcement through directed paths of length two, confirming that local network topology captures most of the predictive signal. The corresponding dependence functions further reveal that this effect is smooth, monotonic, and consistent across the full range of values, highlighting its role as a fundamental mechanism shaping social interactions.

At the same time, we identify a small but meaningful subset of links for which individual-level traits dominate the prediction. Although these cases represent only a minor fraction of the data, their analysis reveals that they do not arise from fundamentally different structural conditions. Instead, they correspond to links that are less strongly embedded in the network, where structural signals are weaker and individual predispositions can exert a greater relative influence.

A more detailed characterisation of this subset shows that these links are qualitatively different in their social and structural properties. In particular, they are more likely to correspond to negative or weak ties, exhibit a broader distribution of balanced configurations, and are less consistently embedded in structurally coherent triads. In contrast, links dominated by triadic influence are strongly associated with positive relationships, higher levels of structural balance, and a pronounced tendency to appear in fully or nearly fully balanced configurations. These findings indicate that structural mechanisms are not only more predictive, but also more closely aligned with stable and socially reinforced relationships.

Importantly, our analysis also clarifies the role of links without indirect structural paths. Contrary to previous interpretations, we find that these links are not explained by individual traits, but rather by the absence of structural reinforcement itself. Even when triadic influence is zero, it remains the dominant explanatory feature, suggesting that the model captures the lack of embeddedness as an informative signal.

Taken together, these results support a nuanced view of social tie formation. Rather than a simple dichotomy between individual and structural drivers, the network exhibits a layered organisation in which structural mechanisms dominate globally, while individual-level effects emerge in specific, less constrained contexts. This perspective is consistent with a dynamic interpretation in which early or weakly embedded ties are more sensitive to individual traits, whereas more established relationships become increasingly governed by structural reinforcement and balance.

More broadly, this work highlights the value of explainable machine learning in network analysis. By combining predictive performance with interpretability, it becomes possible to move beyond accuracy and uncover the mechanisms underlying social behaviour. Future work could extend this approach by incorporating temporal information, allowing for a direct investigation of how individual-driven ties evolve into structurally embedded relationships over time.

\section*{Data and code availability}
The data for this study can be accessed directly from \cite{ruiz2023triadic}. The code developed is publicly available and accessible via the following repository: \url{https://github.com/mpereda/FriendshipFormation}.

%
%

\end{document}